\documentclass[final]{aipproc}

\layoutstyle{8x11single}

\begin{document}

\title{$W$ and $Z$ Production in the Forward Region with LHCb}


\author{Philip Ilten\\ on behalf of the LHCb Collaboration}{
  address={Department of Physics, University College Dublin}}

\begin{abstract}
  Results for $W$ and $Z$ boson production from $pp$ collisions at
  $\sqrt{s} = 7$ TeV in the LHCb experiment are presented. Due to
  LHCb's unique forward acceptance in pseudorapidity of $2.0 \leq \eta
  \leq 4.5$ these results are a test of the Standard Model in the
  forward region, and can be used to better constrain parton density
  functions in the low $x$ kinematic regime.
\end{abstract}

\maketitle

\section{Introduction}

The LHCb detector is a fully instrumented forward arm spectrometer at
the LHC, purpose built for $B$-hadron physics \cite{lhcb2008}. Due to
the forward acceptance of the detector, $2.0 \leq \eta \leq 4.5$, LHCb
provides unique and complimentary precision electroweak measurements
to the pseudorapidity range of the CMS and ATLAS detectors, $| \eta |
< 2.5$.

Partonic cross sections for $W$ and $Z$ production can be calculated
using NNLO electroweak theory to a precision of one percent. However,
PDF distributions introduce an additional uncertainty in the
observable hadronic cross sections at the LHC. The hadronic
uncertainty is dependent upon the rapidity of the electroweak boson,
increasing for larger rapidities. For $y < 2$ an uncertainty of
$\approx 1\%$ is introduced, whereas for $y \approx 5$ the uncertainty
increases to $\approx 8\%$ \cite{mstw}.  Additionally, the expected
sign change in $W^\pm$ charge asymmetry falls inside LHCb
acceptance. Within this review the $Z$ and $W$ cross sections measured
at LHCb are summarised for $W \rightarrow \mu$ and $Z \rightarrow
\mu\mu$ with $37.1 \pm 1.3$ pb$^{-1}$ of data, $Z \rightarrow \tau\tau
\rightarrow \mu e$ with $247.9 \pm 12.7$ pb$^{-1}$ of data, and $Z
\rightarrow \tau\tau \rightarrow \mu e$ with $246.4 \pm 12.6$
pb$^{-1}$ of data (neutrinos are omitted in the notation for brevity)
\cite{wz2011, tau2011}.

\section{$W$ Selection}\label{sec:wsel}

For the $W \rightarrow \mu$ selection a single muon with $p_T > 20$
GeV and $2.0 \leq \eta \leq 4.5$ is used. Two light flavour QCD
backgrounds are considered, decay in flight and punch through of pions
and kaons. These backgrounds are reduced by requiring the summed $p_T$
of all tracks and photons within a cone of $\Delta R \equiv
\sqrt{\Delta \phi^2 + \Delta \eta^2} < 0.5$ to be less than $2$ GeV
and the sum of the associated electromagnetic and hadronic calorimeter
energies over the track momentum to be less than $0.04$. Requiring the
muon impact parameter to be less than $40$ $\mu$m minimises heavy
flavour QCD backgrounds and requiring no additional muons with $p_T >
5$ GeV in the event suppresses electroweak backgrounds.

The signal and background composition of the observed events is
determined using a template fit where the $W \rightarrow \mu$ and QCD
decay in flight template shapes are allowed to float. Electroweak
background shapes are calculated using POWHEG \cite{powheg} and Pythia
\cite{pythia} and normalised to the observed cross sections. The QCD
and heavy flavour shapes are determined from data and, excepting QCD
decay in flight, normalised using simulation. The final template fit
is performed over lepton $p_T$ and shown in Figure \ref{fig:shapes}a.
Further details on background estimation can be found in Reference
\cite{wz2011}.

\section{$Z$ Selection}\label{sec:zsel}

\begin{figure}
  \begin{tabular}{cc}
    \includegraphics[width=0.4\columnwidth]{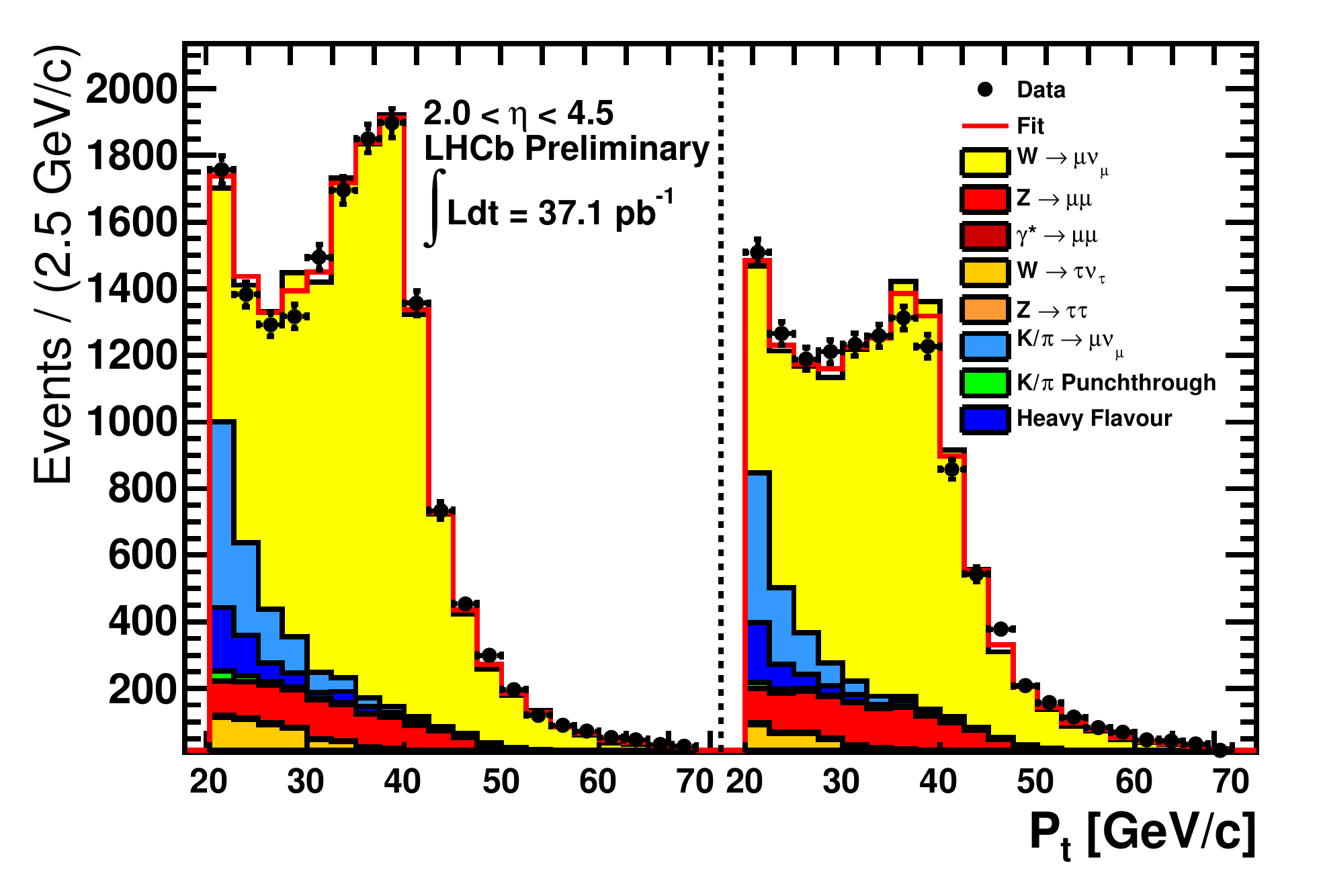} &
    \includegraphics[width=0.4\columnwidth]{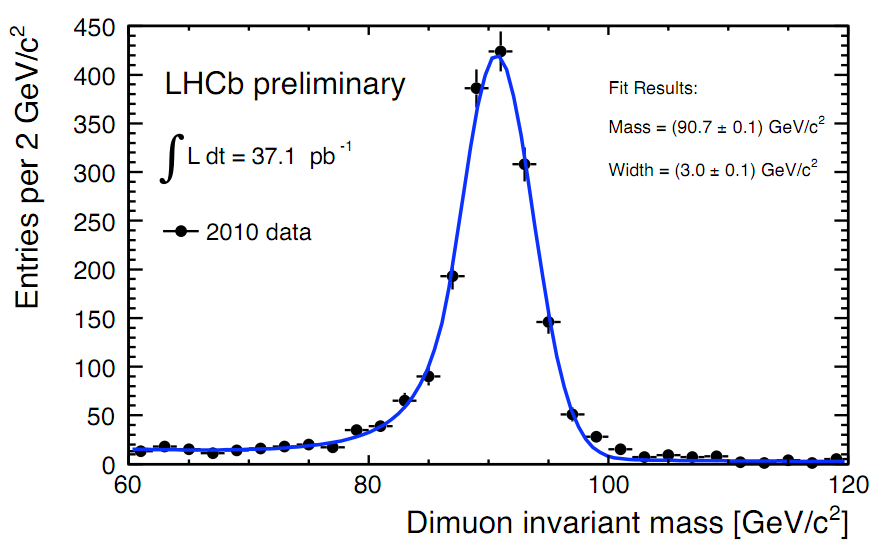} \\ 
    (a) & (b) \\
    \includegraphics[width=0.4\columnwidth]{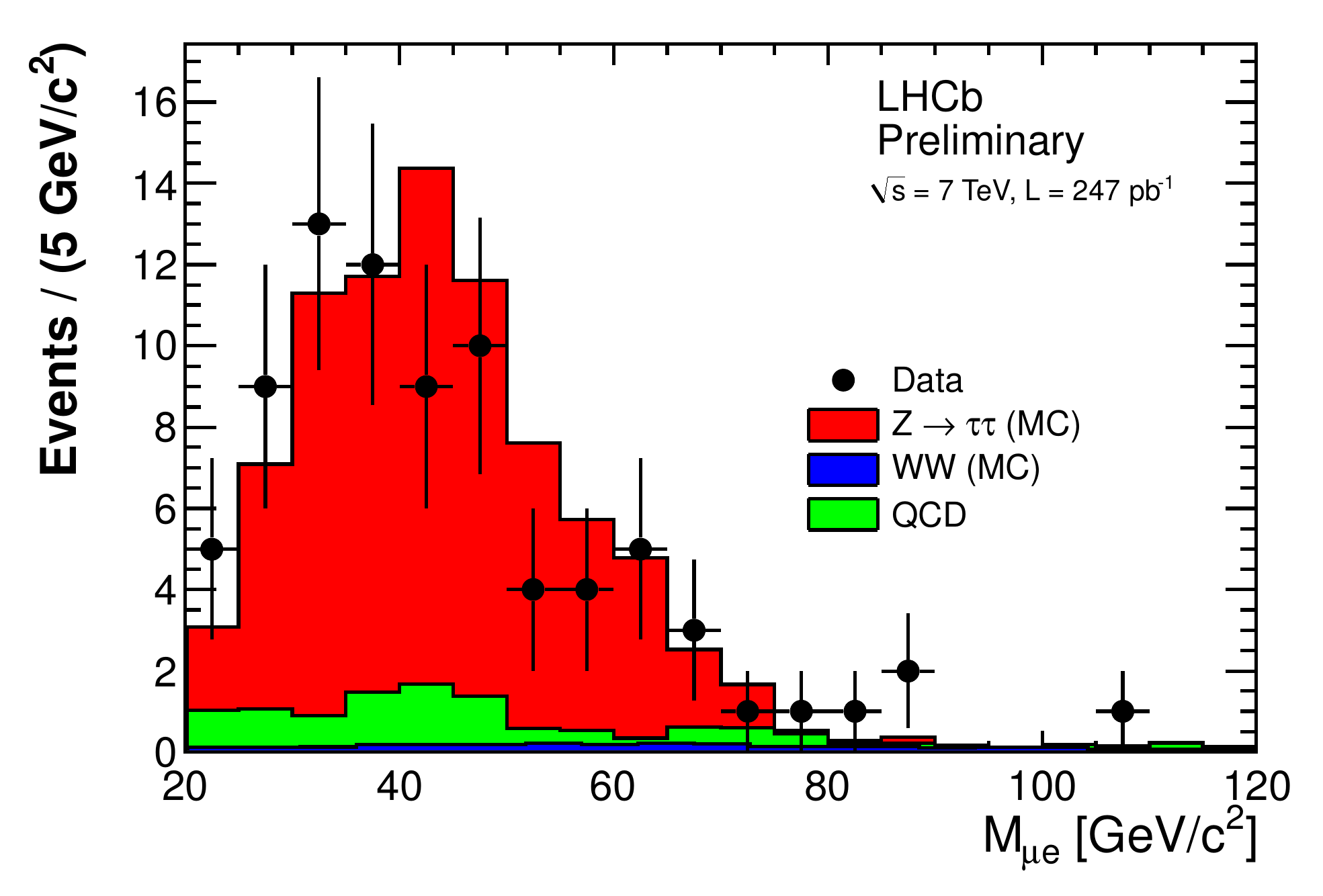} &
    \includegraphics[width=0.4\columnwidth]{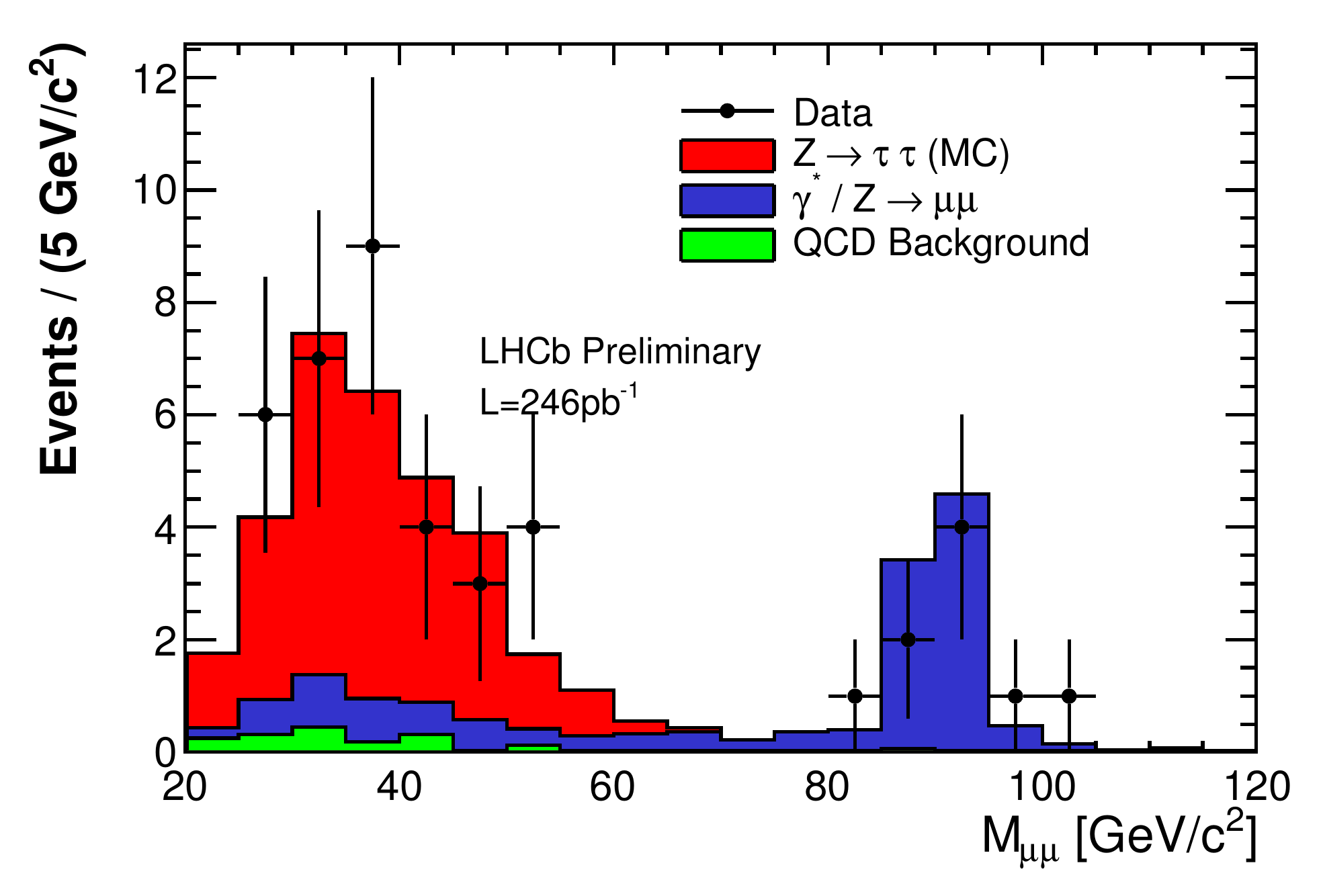} \\
    (c) & (d) \\
  \end{tabular}
  \caption{Transverse momentum distribution for the $W^+$ (left) and
    $W^-$ (right) (a), and the dilepton invariant masses for $Z
    \rightarrow \mu\mu$ (b), $Z \rightarrow \tau\tau \rightarrow \mu
    e$ (c), and $Z \rightarrow \tau\tau \rightarrow \mu\mu$
    (d).\label{fig:shapes}}
\end{figure}

The $Z \rightarrow \mu\mu$ selection requires two opposite sign muons
of $p_T > 20$ GeV with $2.0 \leq \eta \leq 4.5$ and an invariant mass,
shown in Figure \ref{fig:shapes}b, between $60 \leq M_{\mu\mu} \leq
120$ GeV. The QCD background is estimated from same sign events, while
the heavy flavour background shape is determined from data and
normalised using simulation. The $Z \rightarrow \tau\tau$ background
is taken from simulation.

For the $Z \rightarrow \tau\tau \rightarrow \mu e$ and $Z \rightarrow
\tau\tau \rightarrow \mu\mu$ selections a muon with $p_T > 20$ and
opposite sign electron (or muon) with $p_T > 5$ GeV are required in
$2.0 \leq \eta \leq 4.5$ with dilepton invariant mass, Figures
\ref{fig:shapes}c and \ref{fig:shapes}d, greater than $20$ GeV. The
$p_T$ of all tracks within a cone of $\Delta R < 0.5$ around the
leptons is summed and the minimum isolation asymmetry of the two
leptons, $(p_\ell - \sum p _\mathrm{track}) / (p_\ell + \sum p
_\mathrm{track})$, must be greater than $0.8$ to reduce QCD
background. The $t\bar{t}$ and $WW$ backgrounds are suppressed by an
acoplanarity cut on the two leptons, $\Delta \phi > 2.7$. For the $Z
\rightarrow \tau\tau \rightarrow \mu\mu$ channel, cuts on the summed
impact parameter significance of the two leptons, $\sum\mathrm{IP} >
4$, the $p_T$ asymmetry of the two muons, $(p_T^{\mu_1} - p_T^{\mu_2})
/ (p_T^{\mu_1} + p_T^{\mu_2}) > 0.2$, and the dilepton invariant mass,
$M_{\mu\mu} < 80$ GeV, reduce the $\gamma^*/Z \rightarrow \mu\mu$
background.

The QCD and $\gamma^*/Z \rightarrow$ backgrounds are estimated from
data, while the $t\bar{t}$ and $WW$ backgrounds are calculated using
simulation. Further details on background estimation are available in
Reference \cite{tau2011}.

\section{Cross Sections}\label{sec:sigma}

The $W$ and $Z$ cross sections are defined as $\sigma(2.0 \leq
\eta^{\mu, \ell} \leq 4.5, ~ p_T^{\mu, \ell} > 20 ~ \mathrm{GeV}) = (N
- N_\mathrm{bkg}) / (A \epsilon_\mathrm{tot} \mathcal{L} BR)$, where $N$
is total number of events, $N_\mathrm{bkg}$ is number of background
events, $A$ is acceptance, $\mathcal{L}$ is integrated luminosity,
$BR$ is the branching ratio for the process, and
$\epsilon_\mathrm{tot}$ is the total efficiency. The $Z$ total
efficiency is split into $ \epsilon_\mathrm{tot}^Z = A^Z
\epsilon_\mathrm{trk}^\mu \epsilon_\mathrm{trk}^\ell
\epsilon_\mathrm{id}^\mu \epsilon_\mathrm{id}^\ell
\epsilon_\mathrm{sel}^Z$, while the $W$ total efficiency is split into
$\epsilon_\mathrm{tot}^W = A^W \epsilon_\mathrm{trg}^\mu
\epsilon_\mathrm{id}^\mu \epsilon_\mathrm{sel}^W$. The trigger
efficiencies are calculated from the $Z \rightarrow \mu\mu$ data. The
muon track and identification efficiencies are calculated using a tag
and probe method on the $Z \rightarrow \mu\mu$ data while the electron
identification efficiency uses a $Z \rightarrow e e$ data sample. The
electron track efficiency is estimated from simulation and data. The
$Z \rightarrow \mu\mu$ selection efficiency is unity by definition and
the $W \rightarrow \mu$ selection efficiency is found from $Z
\rightarrow \mu\mu$ events. Both the $Z \rightarrow \tau\tau
\rightarrow \mu e$ and $Z \rightarrow \tau\tau \rightarrow \mu\mu$
selection efficiencies are calculated from simulation and data. The
uncertainties for efficiencies from data are statistical, while the
data and simulation efficiency uncertainties are taken as the
difference between simulation and data.

\begin{figure}
  \begin{tabular}{cc}
    \includegraphics[width=0.4\columnwidth]{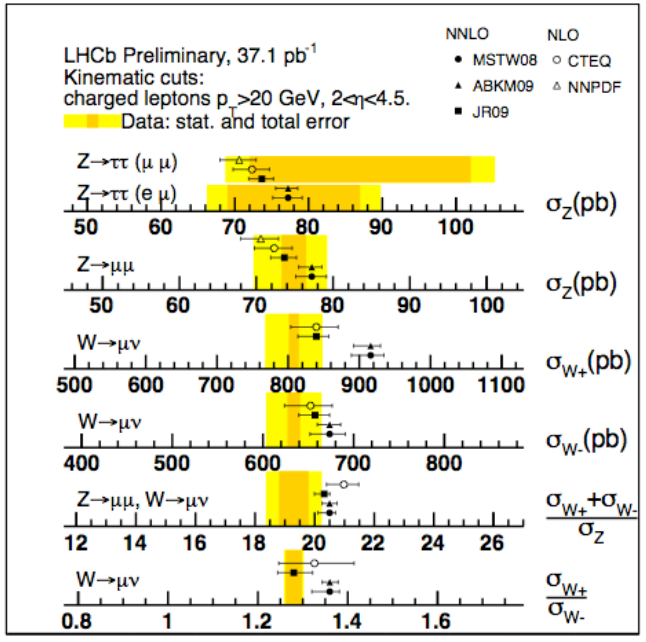} &
    \includegraphics[width=0.5\columnwidth]{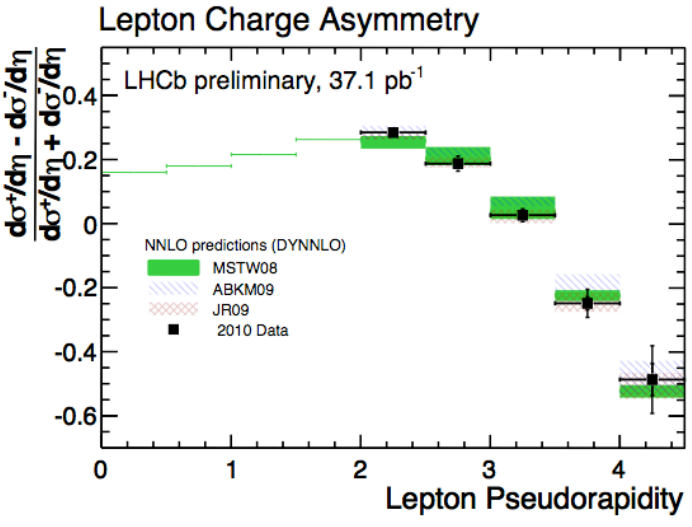} \\
    (a) & (b) \\
  \end{tabular}
  \caption{A summary of the total cross section measurements (a) and
    the measured $W\pm$ charge asymmetry (b).\label{fig:sigma}}
\end{figure}

The $Z \rightarrow \mu\mu$ and $W \rightarrow \mu$ acceptances are
defined as unity, while the $Z \rightarrow \tau\tau$ acceptances are
calculated using both Pythia and Herwig++ \cite{herwig++} with
uncertainty estimated from the difference. The observed cross sections
corrected for FSR \cite{horace} are,
\begin{equation}
  \begin{array}{c}
  \sigma_{W^+ \rightarrow \mu^+} = 808 \pm 7 \pm 28 \pm 28, \quad
  \sigma_{W^- \rightarrow \mu^-} = 634 \pm 7 \pm 21 \pm 22, \quad \sigma_{Z
    \rightarrow \mu\mu} = 74.9 \pm 1.6 \pm 3.8 \pm 2.6 \\
  \sigma_{Z \rightarrow \tau\tau \rightarrow  \mu\mu} = 89 \pm 15 \pm
  10 \pm 5, \quad \sigma_{Z \rightarrow \tau\tau \rightarrow \mu e} =
  79 \pm 9 \pm 8 \pm 4 \\
  \end{array}
\end{equation}
in pb, where the first uncertainty is statistical, the second
uncertainty is systematic estimated from background, efficiency, and
acceptance uncertainty, and the third is luminosity uncertainty. A
comparison of the cross sections to NLO theory from MCFM \cite{mcfm}
and FEWZ \cite{fewz} using the CTEQ \cite{cteq} and NNPDF \cite{nnpdf}
PDF sets and NNLO theory from DYNNLO \cite{dynnlo} using the MSTW08
\cite{mstw}, ABKM09 \cite{abkm}, and JR09 \cite{jr} PDF sets is given
in Figure \ref{fig:sigma}a while the differential $W^\pm$ asymmetry is
given in Figure \ref{fig:sigma}b.

\section{Conclusion}

Cross sections in the forward region, $2.0 \leq \eta \leq 4.5$, have
been presented for $W \rightarrow \mu$, $Z \rightarrow \mu\mu$, $Z
\rightarrow \tau\tau \rightarrow \mu \mu$, and $Z \rightarrow \tau\tau
\rightarrow \mu e$ processes and agree well with NLO predictions. The
differential $W^\pm$ charge asymmetry has been measured and matches
NLO and NNLO predictions. The continuation of these analyses will help
further constrain PDF's and reduce their uncertainties at low $x$.

\begin{theacknowledgments}
  Copyright CERN for the benefit of the LHCb Collaboration.
\end{theacknowledgments}

\bibliographystyle{aipproc}

\bibliography{531_Ilten}

\end{document}